\documentclass[11pt]{article}
\usepackage{amssymb, amsmath}

\makeatletter
\@addtoreset{equation}{section}

\makeatother

\newcommand{\qed}{{\vrule width8pt height8pt}}

\renewcommand{\L}{\mathcal{L}}
\newcommand{\Z}{\mathbf{z}}

\newtheorem{thm}{Theorem}
\newtheorem{theorem}{Theorem}
\newtheorem{dfn}[thm]{Definition}
\newtheorem{prop}[thm]{Proposition}
\newtheorem{lemma}[thm]
{Lemma}

\begin{document}
\pagestyle{empty}

\title{AN OPTIMAL CONTROL FORMULATION FOR
INVISCID INCOMPRESSIBLE IDEAL FLUID FLOW}

\author{Anthony M. Bloch\thanks{
  Research partially supported by the NSF
 and AFOSR.
}
  \\Department of Mathematics
  \\ University of Michigan \\ Ann Arbor MI 48109
  \\{\small abloch@math.lsa.umich.edu}
  \and
Peter E. Crouch\thanks{
Work supported in part by NSF and NATO.}
  \\ Center for Systems Science
  \\and Engineering
  \\ Arizona State University
  \\Tempe, AZ 85287
  \\{\small peter.crouch@asu.edu}
\and
Darryl D. Holm\thanks{Work supported by DOE
, under contract W-7405-ENG-36.}
\\Mathematical Modeling and Analysis
\\ Theoretical Division
\\ Los Alamos National Laboratory\\
Los Alamos, NM 87545\\
{\small dholm@lanl.gov}
\and
Jerrold E. Marsden\thanks{
Research partially supported by NSF and AFOSR.}
\\Control and Dynamical Systems 107-81
\\ California Institute of Technology
\\ Pasadena, CA 91125
\\ {\small marsden@cds.caltech.edu}
\\
\\{\date PProc. of the 39th IEEEE Conference on Decision and
Control}
\\
{\date SSydney, Australia, December 2000}
\\
{\date LLA-UR-00-5005}
}

\maketitle

\begin{abstract}
In this paper we consider the Hamiltonian formulation of the equations of
incompressible ideal fluid flow from the point of view of optimal control
theory. The equations
are compared to the finite symmetric rigid body equations analyzed
earlier by the authors.
We discuss various aspects of the Hamiltonian structure of the Euler
equations and show in particular that the optimal control approach leads to
a standard formulation of the Euler equations --
the so-called impulse equations in their Lagrangian form.
We discuss various other aspects of the Euler equations from a
pedagogical point of view.
We show that
the Hamiltonian in the maximum principle is given by the pairing of the
Eulerian impulse density with the velocity. We provide a comparative
discussion of the flow equations in their Eulerian and Lagrangian form and
describe how these forms occur naturally in the context of optimal control.
We demonstrate that the extremal equations corresponding to the optimal
control problem for the flow have a natural canonical symplectic structure.
\end{abstract}
\section{Introduction}
In this paper we consider the Hamiltonian formulation of the equations of
incompressible ideal fluid flow from the point of view of optimal control
theory. Our goal in this work is to compare the fluid
equation arising in this fashion with the symmetric generalized rigid body
equations derived in Bloch and Crouch [1996], Bloch, Brockett and Crouch
[1997]
and Bloch, Crouch, Marsden and Ratiu [1998]. In these papers we
showed that a natural control approach led to a form
of the rigid body equations on
$\operatorname{SO}(n)\times \operatorname{SO}(n)$ rather
than on $T^*\operatorname{SO}(n)$.
In contrast here we show that the optimal control approach leads to
a standard formulation of the Euler equations --
the so-called impulse equations in their Lagrangian form. A nice
survey of the impulse equations in the various forms can be found
in Russo and Smereka [1999] (see also Kuz'min [1983], Osledets [1989]
and Maddocks and Pego [1995] for related work). Lie-Poisson reduction is an
important tool in rationalizing many of these approaches (see
e.g. Marsden and Weinstein [1984]). In particular, the Hodge projection
to the divergence free vector fields, is a Poisson map since it may be
regarded as the dual of the natural inclusion (this is a standard
result; see Marsden and Ratiu [1994]). Thus, the Hodge projection naturally
takes the unconstrained Poisson system to the constrained one. Russo and
Smereka and others concern themselves with the numerical aspects of these
equations, something we do not consider here.

These impulse equations are not symmetric in the same
sense as the symmetric rigid body equations -- i.e. we do not obtain
two symmetric equations evolving on two copies of the diffeomorphism
group, but it is possible to get a more symmetric formulation which
we intend to discuss in a forthcoming publication
 and that we mention briefly in the
conclusions. 

In the remainder of the introduction we recall the standard and
symmetric rigid body equations.

We recall from Manakov [1976] and Ratiu [1980] that the left
invariant generalized rigid body equations on
$\operatorname{SO}(n)$ may be written as
\begin{align}
\dot Q&= Q\Omega  \nonumber \\
\dot M&= [M,\Omega]\,, \tag{RBn}
\label{rbl}
\end{align}
where $Q\in \operatorname{SO}(n)$ denotes the configuration space
variable (the attitude of the body), $\Omega=Q^{-1}\dot{Q} \in
{so}(n)$ is the body angular velocity, and
\[
M:=J(\Omega)=\Lambda\Omega +\Omega\Lambda \in
{so}(n)
\]
          is the body angular momentum. Here
$J: {so}(n) \rightarrow  {so}(n) $ is the symmetric
(with respect to the above inner product) positive definite operator
defined by
\[
J(\Omega)=\Lambda\Omega +\Omega\Lambda ,
\]
          where $\Lambda$ is
a diagonal matrix satisfying $\Lambda_i + \Lambda_j >0$ for
all $i \neq j$. For $n=3$ the elements of $\Lambda_i$
are related to the standard diagonal moment of inertia tensor $I$ by
$I_1 = \Lambda_2 + \Lambda_3$,  $I_2 = \Lambda_3 + \Lambda_1$,
          $I_3 = \Lambda_1 + \Lambda_2$.

The equations $ \dot{ M } =  [ M, \Omega
] $ are readily checked to be the Euler-Poincar\'e equations on
${so}(n)$ for the Lagrangian
\[
l ( \Omega ) = \frac{1}{2}  \left\langle  \Omega , J
( \Omega )
\right\rangle .
\]

The left invariant
symmetric rigid body system is given by
the first order equations
\begin{align}
\dot Q&= Q\Omega \nonumber \\
\dot P&= P\Omega
\label{rbnl}
\end{align}
where $\Omega$ is regarded as a function of $Q$ and $P$ via the
equations
\[
\Omega :=J^{-1}(M)
\in {so}(n)
\quad \mbox{and} \quad M := Q^TP-P^TQ.
\]

These equations can be derived from the following optimal control
problem:

\begin{dfn}\label{rboptcontprob}
          Let $T >0 $, $Q _0, Q _T \in \operatorname{SO}(n)$
be given and fixed. Let the rigid body optimal control problem be given by
\begin{equation}
\mathop{\rm min}_{U\in
{so}(n)}\frac{1}{4}\int_0^T
\langle U,J(U)\rangle dt
\label{optr}
\end{equation}
subject to the constraint on $U$ that there be a curve
$Q (t) \in \operatorname{SO}(n)$ such that
\begin{equation}
\dot Q=QU\qquad Q(0)=Q_0,\qquad Q(T)=Q_T.
\label{eqnr}
\end{equation}

\end{dfn}

\begin{prop} The rigid body optimal control problem \ref{rboptcontprob}
has extremal evolution equations
(\ref{rbnl}) where $P$ is the costate vector given by the maximum
principle.

The optimal controls in this case are given by
\begin{equation}
U=J^{-1}(Q^TP-P^TQ).
\end{equation}
\end{prop}

In \S\ref{optsec} we  derive the impulse equations
for fluid flow from the optimal control point of view.

\section{Inviscid, Incompressible, Fluid Flow}
In this section we introduce the usual dynamics for inviscid, incompressible
fluid flow, impulse density and the vorticity dynamics. The basic equations
we consider are:
\begin{equation}\label{eq1}
\frac{\partial v}{\partial
t}+(v\cdot\operatorname{grad})v=-\operatorname{grad}p;\quad
\operatorname{div}v=0
\end{equation}
\[
x\in\Omega;\qquad v=v(x,t),\quad p=p(x,t).
\]
We assume, for simplicty only that the flow is in all of space or in a
periodic box so we do not need to deal with boundary conditions. This is
not an essential restriction.

Here, $v$ is the fluid velocity
and $p$ is the pressure. We introduce the impulse
density $z$,
\begin{equation}\label{eq2}
z=v+\operatorname{grad}k.
\end{equation}
where $k$ is an arbitrary scalar field, $k=k(x,t)$. Notice that the
preceding equation gives the (Helmholtz)-Hodge decomposition of $z$. In
other words, the projection of
$z$ to $v$ is the Hodge projection of $z$. We return to this important
remark in the conclusions to gain deeper insight into what is going on
with the calculations to follow. 

Take the time derivative of (\ref{eq2}) to get
\begin{equation}\label{eq3}
\frac{\partial z}{\partial
t}-v\times\operatorname{curl}
z=\operatorname{grad}\Lambda,\quad\operatorname{div}v=0
\end{equation}
where
\[
\Lambda=\frac{\partial k}{\partial t}-p-\frac12v\cdot v;
\]
$\Lambda$ is called the gauge. Any choice of gauge is possible, but to be
concrete, we consider 
the ``geometric gauge'' $\Lambda=-v\cdot z$. 

With this choice
\[
\frac{\partial z}{\partial
t}+(v\cdot\operatorname{grad})z+(\operatorname{grad}v)^Tz=0,
\quad\operatorname{div}v=0
\]
and $k$ is now fixed by the equation
\[
\frac{dk}{dt}=p-\frac12v\cdot v.
\]
Let $\Z=\sum_iz_idx_i(=z\cdot dx)$ be the one form corresponding to $z$.
Then
one can easily show that
\[
\L_{\textrm{v}}\Z=(z_*v+v^T_*z)\cdot dx.
\]
Thus the impulse density is governed by:
\begin{equation}\label{eq4}
\frac{\partial \Z}{\partial t}+\L_{\textrm{v}}\Z=0,\quad
\operatorname{div}v=0.
\end{equation}
Hence by applying the exterior differential operator we obtain
\[
\frac{\partial}{\partial
t}d\Z+\L_{\textrm{v}}d\Z=0,\quad\operatorname{div}v=0.
\]

\refstepcounter{theorem}
\begin{lemma}\label{L:1}
$w=\operatorname{curl} z=\operatorname{curl} v$ satisfies the vorticity
equation:
\begin{equation}\label{eq5}
\frac{\partial w}{\partial t}+[v,w]=0
\end{equation}
\end{lemma}

\noindent
\textbf{Proof\quad }$d\Z=\sum_idz_i\wedge
dx_i=\sum_{ij}(\hat{\operatorname{curl} }z)_{ij}dx_i\otimes dx_j$,
where $\hat ab=a\times
b$. We may now compute:
\[
\L_{\textrm{v}}d\Z=\sum_{ij}\Big[v,{\operatorname{curl}
}z\Big]^{\wedge}_{ij}dx_i\otimes dx_j.\qquad
\qed
\]

 We now
quickly review the two coordinate systems associated with the fluid system.
We denote the Lagrange or material variables by
$X_i$ and the Euler or spatial variables by $x_i$, and set
\[
x_i=\phi_i(X,t),\quad 1\le i\le3.
\]
We assume $\phi:\Omega\to\Omega$ is a volume preserving diffeomorphism,
with Jacobian equal to unity, $|\phi_*|=1$.

Let $v(x,t)=$ spatial velocity, so that
\[
\frac{\partial x_i}{\partial t}=v_i(x,t).
\]
Thus $V(X,t)=v(\phi(X,t))$ is the material velocity. Hence
\[
\frac{\partial\phi}{\partial t}(X,t)=v(\phi(X,t))
\]
or
\begin{equation}\label{eq6}
\frac{\partial\phi}{\partial t}=v\circ\phi.
\end{equation}
We note the ``right invariance'' of this system and its evolution on
the ``Group'' $G =\mbox{Diff}_{\mbox{\small vol}}(\Omega)$
of volume preserving diffeomorphisms
of $\Omega$. Setting
\[
\langle a,b\rangle=\int_{\mathbf{R} ^3}a^T(x,t)b(x,t)dx,
\]
and using $|\phi_*|=1$, we obtain the identity
\[
\langle a\circ\phi,b\rangle=\langle a,b\circ\phi^{-1}\rangle.
\]

With this introduction we may introduce the total vorticity equations:
\begin{equation}\label{eq7}
\frac{\partial\phi}{\partial t}=v\circ\phi;\qquad \frac{\partial w}{\partial
t}=[w,v]:\operatorname{div} v=0.
\end{equation}
It is interesting to compare these equations with the right invariant Euler
equations
for the rigid body:
\begin{equation}\label{eq8}
\dot Q=\Omega Q;\qquad\dot M=[\Omega,M]
\end{equation}
\[
[\Omega, M]=\Omega M-M\Omega\left(\begin{array}{c}
=[M,\Omega]\mbox{ interpreted}\\
\mbox{as vector fields}\end{array}\right).
\]
These Euler equations are Hamiltonian on $T^*\mbox{SO}(3)$, with the
canonical
symplectic structure. The equivalent statements about (\ref{eq7}) have been
well
studied, (see references in the introduction). However, the derivation of
the symmetric version as in (\ref{rbnl}) provides  our motivation for this
new study.

\section{Optimal Control Problem}\label{optsec}
In this section we introduce an optimal control problem and discuss the
corresponding extremals. The problem can be posed as:
\[
\min_{v(\cdot)}\frac12\int^T_0\langle v,v\rangle dt
\]
subject to:
\begin{equation}\label{eq9}
\operatorname{div}v=0;\qquad \frac{\partial\phi}{\partial t}=v\circ\phi
\end{equation}
and
\[
\phi(X,0)=\phi_0(X),\qquad \phi(X,T)=\phi_T(X)\mbox{ fixed},
\]
and, for flow in all of space, suitable conditions at infinity.

This optimal control
problem is of course identical to the standard Hamilton principle
for ideal fluid mechanics. However our goal here is to analyze
it from the point of view of the Pontryagin maximum principle.

We solve this problem by introducing Lagrange multipliers
and the cost
\[
J(v,\phi,\pi,k)=\int^T_0\bigg(\bigg\langle
\pi,v\circ\phi-\frac{\partial\phi}{\partial t}\bigg\rangle-\frac12\langle
v,v\rangle+\langle k,\operatorname{div}v\rangle\bigg)dt
\]
The problem (\ref{eq9}) may be recast as:
$\min J$, subject to $\operatorname{div}v=0$, $\frac{\partial \phi}{\partial
t}=v\circ\phi$, and boundary conditions. 

We may prove the following result:

\refstepcounter{theorem}
\refstepcounter{theorem}
\begin{theorem}\label{T:2}
The extremals of problem {\em(\ref{eq9})} are given by
\begin{align}\label{eq10}
\frac{\partial \pi}{\partial t} =&-(v_*\circ\phi)^T\pi,\quad
\frac{\partial\phi}{\partial t}=v\circ\phi\\
v =&\pi\circ\phi^{-1}-\mbox{\rm grad }k,\quad\mbox{\rm div
}v=0.\qquad\qed\nonumber
\end{align}
\end{theorem}

\medskip
\noindent
\textbf{Sketch Proof}

\begin{eqnarray*}
\delta J&=&\int^T_0(\langle \pi,\delta
v\circ\phi+(v_*\circ\phi)\delta\phi-\frac\partial{\partial
t}\delta\phi\rangle-\langle v,\delta v\rangle\\
&&+\langle k,\mbox{ div }\delta v\rangle)dt\\
&=&\int^T_0(\langle \pi,\delta v\circ\phi\rangle-\langle v,\delta
v\rangle+\langle k,\mbox{ div }\delta v\rangle)dt\\
&&+\int^T_0\langle \pi,(v_*\circ\phi)\delta\phi-\frac{\partial}{\partial
t}\delta\phi\rangle dt
\end{eqnarray*}
Noting that $\delta v(\infty,t)=\delta\phi(x,0)=\delta\phi(x,T)=0$, we
obtain
\begin{eqnarray*}
\delta J&=&\int^T_0\langle \pi\circ\phi^{-1}-v-\mbox{ grad }k,\delta v\rangle
dt\\
&&+\int^T_0\langle(v_*\circ\phi)^T\pi+\frac{\partial \pi}{\partial
t},\delta\phi\rangle dt.
\end{eqnarray*}
The system (\ref{eq10}) follows immediately.\qquad\qed
\bigskip 

We note that the system (\ref{eq10}) should be interpreted in terms of
Lagrange and Euler variables in the form
\begin{align*}
\frac{\partial \pi}{\partial t}(X,t) =&-\left(\frac{\partial v}{\partial
x}(\phi(X,t),t)\right)^T\pi(X,t),\\
\frac{\partial\phi}{\partial t}(X,t) =&v(\phi(X,t)),
\end{align*}
and
\[
v(x,t)=\pi\circ\phi^{-1}(x,t)-\operatorname{grad}k(x,t).
\]

We now study the Hamiltonian for the extremal flow. Employing the maximum
principle we know that the Hamiltonian corresponding to the problem
(\ref{eq9}) is
\[
H(\pi,\phi)=\langle \pi,v\circ\phi\rangle-\frac12\langle v,v\rangle.
\]

 We introduce the vector potential for $v$
\[
v=\operatorname{curl} \psi;\qquad \operatorname{div}\psi=0,
\]
(and $ \psi \rightarrow 0 $ at infinity in all of space). Thus
\[
\omega=\operatorname{curl} v=\mbox{curl curl}\psi=-\Delta\psi+\mbox{grad
div}\psi=-\Delta\psi,
\]
so
\[
\omega=-\Delta\psi;\qquad\Delta=\mbox{Laplacian};.
\]
Thus $\psi=A\omega$ where $A$ is an integral operator. From these
identities we may write:
\begin{eqnarray*}
H(\pi,\phi)&=&\langle \pi\circ\phi^{-1},\mbox{ curl}\psi\rangle-\frac12\langle
v,\mbox{ curl}\psi\rangle\\
&=&\langle\operatorname{curl}
\pi\circ\phi^{-1},\psi\rangle-\frac12\langle\operatorname{curl}
v,\psi\rangle
\end{eqnarray*}
But
\[
v=\pi\circ\phi^{-1}-\operatorname{grad}k
\]
so
\[
\operatorname{curl} v=\operatorname{curl}
\pi\circ\phi^{-1}
\]
and
\begin{eqnarray}\label{eq11}
H(\pi,\phi)&=&\frac12\langle\operatorname{curl}
\pi\circ\phi^{-1},\psi\rangle\\
&=&\frac12\langle
\pi\circ\phi^{-1},v\rangle\\
&=&\frac12\langle
\omega,A\omega\rangle\\
&=&\frac12\langle\operatorname{curl}
\pi\circ\phi^{-1},A\mbox{
curl}\pi\circ\phi^{-1}\rangle.
\end{eqnarray}

We now compute along extremals (\ref{eq10})
\[
\frac{\partial}{\partial
t}\sum_i\pi_i(X,t)d\phi_i(X,t)=0
\]
or
\[
\frac{\partial}{\partial
t}\sum_i\pi_id\phi_i=0.
\]
Hence
\[
\frac{\partial}{\partial
t}\sum_id\pi_i\wedge d\phi_i=0.
\]
Thus the ``canonical two
form''
\[
\sum_id\pi_i\wedge d\phi_i=\sum_{ijk}\frac{\partial
\pi_i}{\partial
X_j}\frac{\partial\phi_i}{\partial X_k}dX_j\wedge
dX_k
\]
is constant along extremals. We now have the following
critical result.

\refstepcounter{thm}
\begin{lemma}\label{L:3}
Let $z=\pi\circ\phi^{-1}$.
Along extremals {\em(\ref{eq10})}
\[
\frac{\partial z}{\partial
t}+z_*v+v^T_*z=0.\qquad\qed
\]
\end{lemma}

Thus $\Z=z\cdot dx$
satisfies
\begin{equation}\label{eq14}
\frac{\partial \Z}{\partial
t}+\L_{\textrm{v}}\Z=0,\qquad
\frac{\partial\phi}{\partial
t}=v\circ\phi
\end{equation}

It follows
that we have recovered the evolution of the impulse density of
the
fluid flow, equation (\ref{eq4}). Note that $H=\frac12\langle
z,v\rangle$, where $-z\cdot v=\Lambda$ is the geometric gauge.
The following result relates $z$ to the canonical two form.

\begin{lemma}\label{L:4}
$\phi^{-1*}\sum_id\pi_i\wedge d\phi_i=\sum_idz_i\wedge
dx_i=d\Z$.\qquad{\vrule width8pt height4pt}
\end{lemma}

\vspace{12pt}
Substituting the relation
\[
z=v+\mbox{\rm grad }k
\]
into the system (\ref{eq14}) recovers the system (\ref{eq1}) where
\begin{equation}\label{eq15}
\frac{dk}{dt}=p-\frac12\mbox{ v.v.}
\end{equation}
However $k$ is not arbitrary, since it is determined by the extremal system
(\ref{eq10}). In fact
\[
\mbox{\rm div }z=\mbox{\rm div }\pi\circ\phi^{-1}=\mbox{\rm div grad }k=\Delta
k.
\]
Thus,
\begin{align}\label{eq16}
k =&A\mbox{\rm div }z=A\mbox{\rm div }\pi\circ\phi^{-1}.\\
v =&\pi\circ\phi^{-1}-\mbox{\rm grad }A\mbox{\rm div
}\pi\circ\phi^{-1}.\nonumber
\end{align}
Hence the pressure $p$ is also determined by the flow.

\begin{lemma}\label{L:5}
By augmenting the cost functional in the optimal control problem
{\em(\ref{eq9})} by a potential function $\eta$
\[
\int^T_0\bigg(\frac12\langle
\mbox{v.v}\rangle-\int_{\Omega}\eta\circ\phi\bigg)dt
\]
the extremal flow satisfies the system {\em(\ref{eq1})} with the pressure
determined by
\[
\frac{dk}{dt}=p-\eta-\frac12\mbox{v.v.}
\]
$k$ and $v$ are determined from {\em(\ref{eq16})}.\qquad\qed
\end{lemma}

Thus any pressure $p$ may be obtained via a suitable potential $\eta$.

\section{Hamiltonian Structure of Extremals}
We now briefly explore the Hamiltonian nature of 
(\ref{eq1}) and
the extremal equations (\ref{eq10}). 
If $u$ is a smooth function of $x$ and $t$, and $h[u]$ is a function of
$x,t$
and the jet of $u$, let
\[
H[u]=\int_{{\mathbf R}^3}h[u]dx.
\]
Define
\[
\delta H[u]=\int_{{\mathbf R}^3}\left\langle\frac{\delta H(u)}{\delta
u},\delta u \right\rangle dx.
\]
We have the following result:

\refstepcounter{theorem}
\refstepcounter{theorem}
\refstepcounter{theorem}
\begin{theorem}\label{T:6}
For the Hamiltonian {\em(3.6)}
\[
\frac{\delta H}{\delta \pi}(\pi,\phi)=v\circ\phi;\qquad\frac{\partial
H}{\partial \phi}(\pi,\phi)=(v_*\circ\phi)^T\pi.\qquad\qed
\]
\end{theorem}

Thus the extremal equations (\ref{eq10}) may be written as
\begin{equation}\label{eq19}
\frac{\partial \pi}{\partial t}=-\frac{\delta H}{\delta\phi};\qquad
\frac{\partial \phi}{\partial t}=\frac{\delta H}{\delta \pi}.
\end{equation}
These equations are canonical with respect to the natural symplectic form on
$L_2(\mathbf{R} ^3:\mathbf{R} ^3)\times L_2(\mathbf{R} ^3:\mathbf{R} ^3)$
\[
\omega((X_1,Y_1),(X_2,Y_2))=\int_{\mathbf{R} ^3}(Y_2\cdot X_1-X_2\cdot
Y_1)dx
\]
Thus we have expressed the extremal equations (\ref{eq10}) in terms of a
canonical Hamiltonian system.

\section{Conclusions}
As described earlier, the Euler (impulse) equations
(\ref{eq10}) are not quite in the symmetric form that we obtain
in the rigid body setting -- i.e. we do not get a symmetric flow on
two copies of the diffeomorphism group. However, it is possible
to extend the analysis to this setting by factoring $\pi$
as
\begin{equation}
\pi=r\circ\psi,
\end{equation}
where
\begin{equation}
\frac{\partial\psi}{\partial t}=v\circ\psi
\end{equation}
and $\psi$ evolves on the diffeomorphim group. Thus we do
get  symmetric equations for $\phi$ and $\psi$ coupled
to an interesting radial equation for $r$. This also has an
analogue in the finite-dimensional setting -- one allows
$P$ to be in $Gl(n)$ and considers the polar decomposition
$P=RK$ where $R$ is symmetric positive definite and $K$ lies
in $\operatorname{SO}(n)$. We shall describe the details of this
analysis in a  forthcoming publication.

In addressing these issues, a deeper understanding of both the
Hamiltonian and variational structure as well as the geometry is needed.
For example, we can obtain more insight into some of the calculations done
in this paper as follows. Consider the Hodge projection
$\mathbf{P}:
{X}
\rightarrow
{X} _{\rm vol}$ taking a vector field $z$ to its divergence
free part parallel to the boundary. As we have mentioned in the
introduction, using the $L _2$ pairing, this map is a Poisson
map. Taking the $L _2$ kinetic energy as the Hamiltonian on the
unconstrained space
${X}$ as well as on the constrained space ${X}_{\rm
vol}$, we conclude that the corresponding Hamiltonian systems with their
Lie Poisson bracket structures are mapped one to the other (including
integral curves) by the Hodge projection. This simple remark is, in fact,
the essense of what is going on in relaxing the divergence free constraints
and in relating the Hamiltonian structure in the formalism of Osledets,
Buttke, and Kusmin. We have, in fact, shown some aspects of this remark
in the above direct calculations. A deeper problem, to which we will
return in other work, is to carry this out in material representation,
where one needs a nonlinear Hodge decomposition, similar to the Moser
decomposition (a diffeomorphism group analogue of the polar
decomposition) discussed in Ebin and Marsden [1970]. Many of these issues
are addressed in work of Brenier; see, eg, Brenier [1999].

{\bf Acknowledgement:} We would like to thank Peter Smereka for
useful conversations.

\section{References}
\begin{description}

\item Benjamin T. Brooke [1984] 
Impulse, Flow Force and Variational Principles,
{\it IMA Journal of Applied Mathematics} {\bf32}, 3-68.

\item Bloch, A.M., R. W. Brockett, and P.E. Crouch [1997]
Double bracket equations and geodesic flows on symmetric
spaces. {\it Comm. Math Phys} {\bf 187},357-373.

\item Bloch, A. M. and P. E. Crouch [1996]
Optimal control and geodesic flows.
{\it Systems and Control Letters}
{\bf 28}, 65-72.

\item Bloch, A. M., P. E. Crouch J.E. Marsden and T.S. Ratiu [1998]
Discrete rigid body dynamics and optimal control
{\it Proceedings of the 37th CDC}, IEEE, 2249-2254.

\item Bloch, A.M., P.S. Krishnaprasad, J.E. Marsden and T.S. Ratiu [1994]
Dissipation induced instabilities, {\it Ann. Inst.
H. Poincar\'e, Analyse Nonlineare} {\bf 11}, 37--90.

\item Brenier, Y. [1999] Minimal geodesics on groups of volume-preserving
maps and generalized solutions of the Euler equations. {\it Comm. Pure
Appl. Math.} {\bf 52}, 411--452. 

\item Buttke, T.F. [1993] Velocity methods: Lagrangian numerical
methods which preserve the Hamiltonian structure of
incompressible fluid flow. In {\it Vortex Flows and Related
Numerical Methods} (ed. J.T. Beale, G.H. Cottet and S.
Hueberson), Kluwer.

\item Buttke, T.F. and A.J. Chorin [1993], Turbulence
calculations in magnetization
variables, {\it Appl. Numer. Maths. }{\bf 12},
47.

\item  Ebin, D.G. and J.E. Marsden [1970] Groups of
diffeomorphisms and the
motion of an incompressible fluid, {\it Ann.
Math.} {\bf 92}, 102-163.

\item Hamel, G. [1904] Die
Lagrange-Eulerschen Gleichungen der Mechanik,
{\it Z. f\"ur
Mathematik u. Physik} {\bf50}, 1-57.

\item Hamel, G. [1949] {\it
Theoretische Mechanik},
Springer-Verlag.

\item Holm, D.D. [1983]
Magnetic tornadoes: three-dimensional affine motions in
ideal
magnetohydrodynamics, {\it Physica D} {\bf 8} (1983), 170-182.

\item
Kuz'min, G.A. [1983] Ideal incompressible hydrodynamics in terms of
the
vortex momentum density, {\it Phys. Let.} {\bf 96A},
88-90.

\item Lamb, H. [1932] {\it Hydrodynamics}, 6th ed., Cambridge
University Press.

\item Maddocks, J.H. and R.L. Pego [1995] An
unconstrained Hamiltonian formulation for
incompressible fluid flow
{\it Comm. Math. Phys.} {\bf 170}, 207-217.

\item Manakov, S.V. [1976]
Note on the integration of Euler's equations of the dynamics of
and $n$-dimensional rigid body.
{\it Funct. Anal. and its Appl.} {\bf 10}, 328--329.

\item  Marsden, J.E.,
and T. Ratiu [1994] {\it Introduction to Mechanics and
Symmetry},
Texts in Applied Math., {\bf17}, Springer-Verlag,
2nd edition, 1999.

\item Marsden, J.E. and A. Weinstein [1983] Coadjoint
orbits, vortices, and
Clebsch variables for incompressible fluids,
{\it Physica D} {\bf7},
305-323.


\item Osledets, V.I. [1989], On a new way of writing the
Navier Stokes equation: The Hamiltonian formalism, 
{\it Russ. Math. Surveys} {\bf 44}, 210-211.

\item Ratiu, T. [1980] The motion of the free
n-dimensional rigid body. {\it Indiana U. Math. J.},
{\bf 29}, 609-627.

\item Russo, G. and P. Smereka [1999]
Impulse formulation of the Euler equation:
general properties and
numerical methods, J. Fluid Mechanics {\bf391}, 189-209.

\item
Serrin, J. [1959] in {\it Mathematical Principles of Classical
Fluid Mechanics},
Vol. VIII/1 of Encyclopedia of Physics, edited by S.
Fl\"ugge (Springer-Verlag, Berlin),
sections 14-15, pp.
125-263.

\item Smereka, P. [1996] A Vlasov description of the Euler
equation,
{\it Nonlinearity}, {\bf 9},
1361.

\end{description}
\end{document}